\documentclass[final,5p,times]{elsarticle}
\usepackage{graphicx}
\usepackage{epstopdf}
\usepackage{tabularx}
\usepackage{amssymb}
\usepackage{amsmath}
\usepackage{diagbox}
\usepackage{color}
\usepackage{balance}
\usepackage{subfigure}
\usepackage{algorithm}
\usepackage{algorithmic}
\graphicspath{{figure/}}
\usepackage{booktabs}
\usepackage{multirow}
\usepackage{CJK}

\allowdisplaybreaks[4]

\allowdisplaybreaks[4]
\journal{Knowledge-Based Systems}
\begin{document}

\begin{frontmatter}

\title{Enhancing multilingual speech recognition in air traffic control by sentence-level language identification
\tnoteref{label1}}
\author[1]{Peng Fan}
\author[1]{Dongyue Guo}

\author[1,2]{JianWei Zhang}
\author[1,2]{Bo Yang}
\author[1,2]{Yi Lin}
\ead{yilin@scu.edu.cn}
\cortext[cor1]{Corresponding author: }
\address[1]{National Key Laboratory of Fundamental Science on Synthetic Vision, Sichuan University, wangjiang road, Chengdu, Sichuan, China}
\address[2]{College of Computer Science, Sichuan University, wangjiang road, Chengdu, Sichuan, China}

\begin{abstract}

Automatic speech recognition (ASR) technique is becoming increasingly popular to improve the efficiency and safety of air traffic control (ATC) operations. However, the conversation between ATC controllers and pilots using multilingual speech brings a great challenge to building high-accuracy ASR systems. In this work, we present a two-stage multilingual ASR framework. The first stage is to train a language identifier (LID), that based on a recurrent neural network (RNN) to obtain sentence language identification in the form of one-hot encoding. The second stage aims to train an RNN-based end-to-end multilingual recognition model that utilizes sentence language features generated by LID to enhance input features. In this work, We introduce Feature-wise Linear Modulation (FiLM) to improve the performance of multilingual ASR by utilizing sentence language identification. Furthermore, we introduce a new sentence language identification learning module called SLIL, which consists of a FiLM layer and a Squeeze-and-Excitation Networks layer. Extensive experiments on the ATCSpeech dataset show that our proposed method outperforms the baseline model. Compared to the vanilla FiLMed backbone model, the proposed multilingual ASR model obtains about 7.50\% character error rate relative performance improvement.
\end{abstract}


\end{frontmatter}

\section{Introduction}
In air traffic control (ATC), radio speech communication is the primary mode of communication between ATC controllers (ATCOs) and pilots \cite{lin_spoken_2021}. 
During the ATC process, ATCOs send control instructions to pilots through radio speech. After receiving the control instructions, pilots repeat the instructions in radio communication to confirm that the instructions are correct. In recent years, automatic speech recognition (ASR) technology has been introduced into the field of ATC to construct intelligent ATC system, which reduces the risk of human error and improve flight safety \cite{lin_spoken_2021,lin_unified_2021,lin_real-time_2020}. After the introduction of ASR technology in the ATC system, it can convert the ATC speech into text and then generate the intention through semantic understanding-related technology. The ATCOs can quickly confirm the correctness of the instruction through the generated text and intention, which greatly reduces the workload of the ATCOs. Therefore, ASR technology is a fundamental part of the new ATC process.

ASR is a well-studied research topic for common applications, which has generated many promising outcomes. However, ASR in ATC poses new challenges and difficulties compared to common ASR research. Our previous work introduced ASR into the ATC safety monitoring framework and converted ATCOs and pilots’ speech into instructions for controlling intent inference \cite{lin_real-time_2020}. In general, the ASR works of the ATC domain are performed based on common ASR approaches, with technical improvements to address the domain-specific characteristics of the ATC speech \cite{juan2020automatic,lin_unified_2021,lin_spoken_2021}. For example, developing ASR systems for ATC poses several challenges, including the lack of transcribed ATC speech data, multilingual speech recognition, and poor speech quality \cite{lin_spoken_2021,yang_atcspeech_2020}.

Incorporating multilingual speech recognition in ATC enhances recognition performance, while recognition failures can negatively impact downstream tasks. This work focuses on multilingual ASR in ATC. Typically, ATCOs and pilots communicate in English, based on the rules published by the International Civil Aviation Organization (ICAO). However, in China, ATCOs and pilots communicate more frequently in Chinese for domestic flights. Therefore, speech on the same frequency on the radio is usually in both Chinese and English, requiring multilingual ASR for the ATC domain \cite{lin_unified_2021}. Multilingual speech usually includes two situations: inter-sentence multilingual and intra-sentence multilingual. As shown in Figure 1, there are two multilingual situations.

\hyphenation{thr-ough}

\begin{enumerate} 
\item[$\bullet$]\textbf{Inter-sentence multilingual}: In China, most pilots of domestic flights are Chinese and communicate with ATCOs in Chinese. However, domestic airlines also employ foreign pilots, and ATCOs and foreign pilots communicate through English. Domestic ATCOs and pilots of foreign international flights naturally communicate through the international standard language of English. Furthermore, due to resource constraints in radio transmission, ATCOs typically communicate with multiple pilots using the same communication frequency. In this case, the ATCO may communicate with a pilot in Chinese and then immediately communicate with the next pilot in English. 

\item[$\bullet$]\textbf{Intra-sentence multilingual}: In practice, the communication between ATCOs and pilots sometimes includes both Chinese and English in one sentence. This is because according to the regulations of the Civil Aviation Administration of China (CAAC) and the ICAO, some professional terms must be expressed in English, and some pilots may use two languages in actual communication. \end{enumerate}

\begin{figure}[h]
	\centerline{\includegraphics[width=0.9\linewidth]{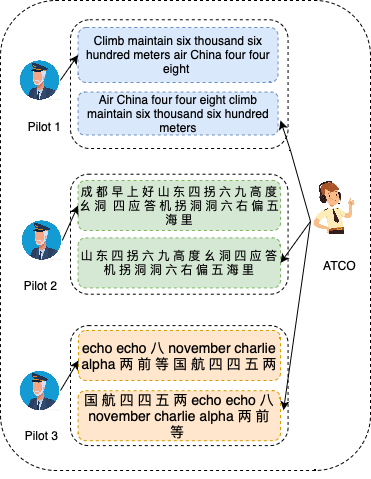}}
	\caption{The description of speech communication in ATC}
\end{figure}





To address the problem of ASR in practical work in ATC, we propose a two-stage multilingual ASR framework. We employ a language identifier (LID) based on recurrent neural networks (RNNs) to obtain sentence language identification as the first stage of the proposed multilingual ASR framework. The second stage of the proposed multilingual ASR framework includes the backbone network and a novel module called SLIL, which is positioned prior to each encoder layer to facilitate the acquisition of sentence language identification by the backbone network. This integration of the SLIL module enhances the overall performance of the multilingual ASR framework.

In the first stage of the proposed methodology, a sentence language classifier based on RNN is trained, followed by optimizing the classifier through Cross-Entropy (CE) loss function during training to ultimately acquire sentence language identification in the form of one-hot encoding.

In the second stage, the backbone network is constructed by cascading the convolutional neural network (CNN) layers, and RNN layers and is jointly optimized by the Connectionist Temporal Classification (CTC) loss function. In this work, we propose a SLIL module, which consists of a Feature-wise Linear Module (FiLM) layer and a Squeeze-and-Excitation Networks (SE) \cite{SE} layer. It can help the acoustic model (AM) learn sentence language identification and improve the performance of the proposed multilingual ASR framework. The FiLM module is proposed to solve image-related questions, the FiLM layer affects neural network computation through simple feature affine transformation based on conditional information. It can be thought of as a generalization of conditional normalization \cite{perez_film_2017}. FiLM module was introduced into speech recognition to improve accurate multi-dialect speech recognition \cite{yoo_highly_2019}. In this work, the sentences are divided into three categories: Chinese, English, and Chinese-English, corresponding to the previously mentioned Inter-sentence multilingual and Intra-sentence multilingual. The sentence language identification in feature-wise transformations makes our encoder more adaptive and able to deal with multilingual more effectively. The SE block adaptively recalibrates channel-wise feature responses by explicitly modeling interdependencies between channels. We introduce the SE network block followed by the FiLM block to form the SLIL module, where the SE block retains the valuable features after FiLM processing and eliminates the useless features.

To sum up, the main contributions of this work are as follows:
\begin{enumerate} 
\item[$\bullet$] We propose a multilingual ASR framework for ATC speech recognition. The proposed framework comprises three main components: a LID model, the FiLM feature generate
network, and an ASR model that utilizes FiLM features.

\item[$\bullet$] We introduce the FiLM module to multilingual ASR for ATC speech recognition by utilizing sentence language identification.
\item[$\bullet$] In this work,  we introduce a new sentence language identification learning module called SLIL that learns sentence language identification and embeds it on the acoustic model (AM) encoder layer. The SLIL module affects neural network computation through simple feature-wise transformation and channel attention based on sentence language identification and improves the speech recognition performance of the model. 
\item[$\bullet$] Extensive experiments demonstrate that our method outperforms previous works on the multilingual ASR task in ATCSpeech dataset.
\end{enumerate}


The remainder of this paper is organized as follows. Section 2 presents a detailed literature review of multilingual ASR approaches and their applications in ATC. Section 3 describes the proposed SLIL language identification adapted block with the AM. Section 4 shows and discusses the experimental results based on the character error rate (CER) metric and compares them with previous works. Section 5 concludes the paper and suggests future directions.

\section{Related work}\label{sec2}

\subsection{Development of ASR}

The classical ASR system is founded upon the utilization of hidden Markov models (HMM) and Gaussian mixture models (GMM) for acoustic and language modeling, as well as search procedures that are governed by the Bayes decision rule \cite{hinton2012deep,sun2017unsupervised}.

In the past few decades, with the development of hardware, deep learning technology has rapidly advanced, and speech recognition technology based on data-driven deep learning has made significant progress. Graves et al. proposed a speech recognition scheme based on RNNs \cite{6638947}. Abdel et al. proposed an automatic speech recognition (ASR) framework based on CNNs \cite{6857341}. Ashish et al. proposed a transformer network based on self-attention to achieve state-of-the-art sequence-to-sequence tasks with extensive data \cite{NIPS2017_3f5ee243}. 

Recently, the end-to-end ASR model has garnered more attention from researchers. The end-to-end ASR model can directly convert the input speech signal to output text without separately training the AM and language model (LM). The standard end-to-end ASR model usually includes several parts. The first is the convolutional layer for preliminary learning of speech features and downsampling of speech signals to shorten the sequence length. Then the encoder learns the high-dimensional acoustic features, and then the decoder converts the acoustic features into text \cite{pmlr-v48-amodei16,8068205}. In previous studies, well-designed Mel-frequency cepstral coefficients (MFCCs) or filterbank (FBANK) features were applied to perform preliminary processing on the raw waveform. In this procedure, the raw speech is divided into frames with a 25 ms frame length and 10 ms shift, and a series of signal processing transformations are applied to convert the 1D waveform into a 2D feature map. Recently, the raw waveform speech is processed by CNNs directly, and the extracted feature map is fed into the neural network for acoustic modeling. This method has achieved state-of-the-art results in many ASR tasks \cite{schneider_wav2vec_2019,fan2021speech}. According to the different decoders of the end-to-end ASR model, this model is divided into Connectionist Temporal Classification (CTC) based model, Attention Encoder-Decoder (AED) based model, and RNN Transducer (RNN-T) model \cite{8068205,graves2006connectionist,pmlr-v48-amodei16,graves2012sequence}. 

\subsection{Multilingual ASR}

Multilingual ASR is a challenging task that aims to recognize speech from different languages using a single model. multilingual ASR has many potential applications in various domains, such as automatic translation, cross-lingual communication, and air traffic control (ATC). In this section, we review some of the recent advances in multilingual ASR based on data-driven deep learning methods.

One of the popular approaches for multilingual ASR is to use an end-to-end (E2E) model that directly maps speech signals to text sequences without any intermediate representations. E2E models can be divided into two categories: sequence-to-sequence (Seq2Seq) models and recurrent neural network transducer (RNN-T) models. Seq2Seq models use an encoder-decoder architecture with an attention mechanism to align the input and output sequences. RNN-T models use a prediction network and a joint network to generate output symbols based on both the encoder output and the previous output symbol.

Toshniwal et al. \cite{toshniwal_multilingual_2018} proposed a Seq2Seq E2E model for multilingual ASR that used graphemes as output units and trained jointly on data from nine different Indian languages without any language identity. They showed that their multilingual model achieved better performance than individual models for each language. Sainath et al. \cite{sainath_two-pass_2019} proposed a two-pass architecture for streaming multilingual ASR that combined an RNN-T decoder and a listen, attend and spell (LAS) decoder that shared an encoder network. They showed that their architecture could improve the performance of streaming multilingual ASR by leveraging both alignment-free and attention-based mechanisms.

Another line of research for multilingual ASR is to use deliberation models that incorporate multiple decoding passes with different objectives or constraints. Hu et al. \cite{hu_deliberation_2020} proposed a deliberation model for multilingual ASR that used two Seq2Seq decoders with different beam sizes and language models to refine the initial hypotheses generated by the first decoder. They showed that their deliberation model could improve the performance of non-streaming multilingual ASR on six languages. Zhang et al.\cite{2022Scaling} extended the deliberation model for streaming multilingual ASR by scaling up the encoder size and using different decoding strategies for different languages.

Language identification is another important factor that affects the performance of multilingual ASR, especially when dealing with low-resource or unseen languages. Language identification can be integrated into E2E multilingual ASR models as an additional task or a feature extractor. Zhang et al. \cite{zhang_streaming_2022} integrated a per-frame Language identification predictor into a cascaded encoder-based RNN-T model and obtained lower character error rates (CERs) than using only one encoder or no Language identification predictor. Waters et al.\cite{waters2019leveraging} trained a streaming Language identification model using RNN-T loss and used it as a feature extractor for E2E ASR models. They showed that their method could improve E2E ASR performance by using Language identification information. 

Moreover, Hou et al.\cite{hou_large-scale_2020} proposed a large-scale joint ASR and Language identification language-independent E2E model hybrid CTC/attention mechanism to solve multilingual ASR tasks. They showed that incorporating language identification could improve the performance of multilingual ASR.

\subsection{Multilingual ASR in ATC}

Several state-of-the-art ASR models were applied to build a benchmark for ATC speech recognition, which was trained on more than 170 hours of ATC speech \cite{juan2020automatic}. Zuluaga et al. \cite{zuluaga2021contextual} integrated contextual knowledge into the ASR model to achieve semi-supervised training, which improved the recognition of callsigns in the ATC instructions. Our previous work \cite{lin_unified_2021} built a unified framework for multilingual speech recognition in the ATC system to translate ATC and pilot multilingual speech-to-text. We introduced transfer learning and pre-training methods to deal with the lack of transcribed speech in the ATC and to improve the performance of the end-to-end deep learning model \cite{2021Improving}. Moreover, we proposed a pronunciation-based sub-word vocabulary containing Chinese and English to solve the problem of vocabulary imbalance in ATC multilingual speech recognition \cite{2021Towards}. In addition, we took into account the complex environment of communication between ATCOs and pilots and applied deep learning methods to learn speech representations from raw waveforms rather than handcrafted features. We fed these representations into the end-to-end ASR model and obtained competitive performance \cite{fan2021speech,lin_atcspeechnet_nodate}. 

In general, research on foreign Air Traffic Control (ATC) speech recognition has primarily focused on English, with limited investigation into multilingual recognition. In China, ATC speech recognition has to consider multilingual speech recognition. Our previous works on multilingual speech recognition at ATC focused on training a powerful end-to-end multilingual recognition model using pre-training and transfer learning, learning speech representations from waveforms using deep learning methods, and designing a pronunciation-oriented vocabulary based on a hybrid Chinese-English vocabulary to improve recognition performance.

 \begin{figure*}[ht!]
 	\centerline{\includegraphics[width=0.95\linewidth]{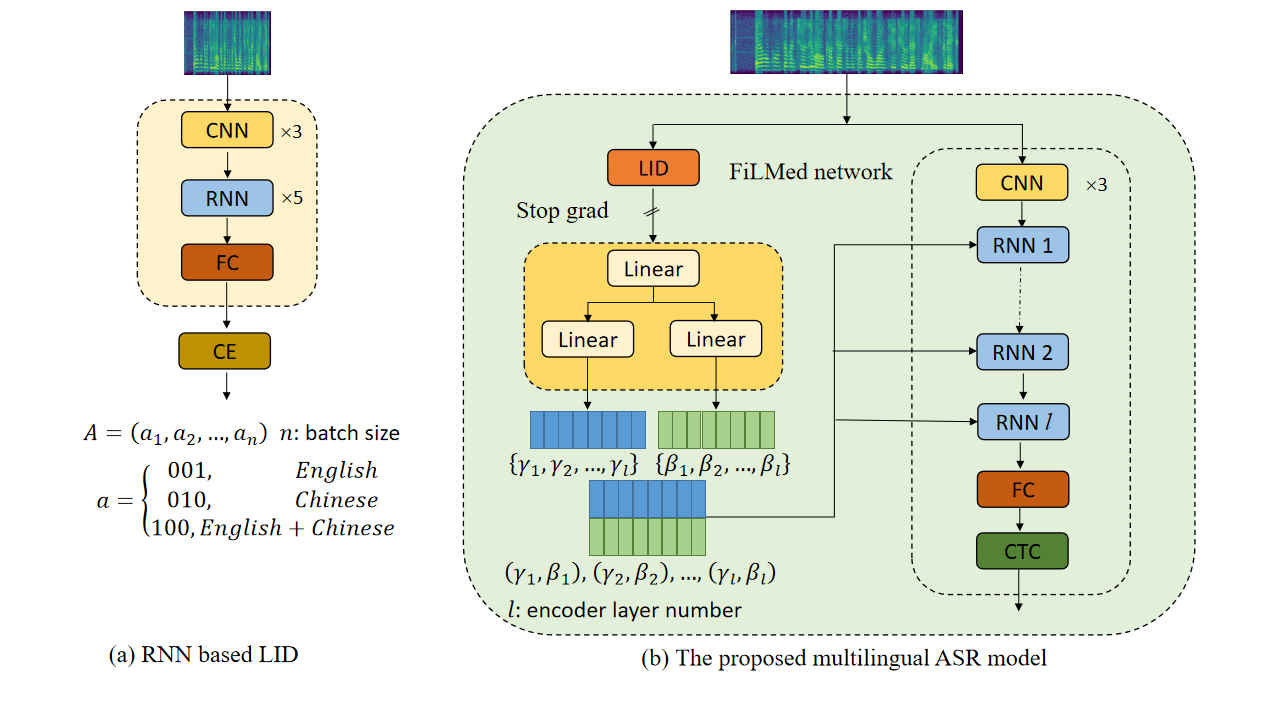}}
 	\caption{The proposed end-to-end multilingual ASR framework has the following overall architecture. (a) The RNN-based LID is trained on ATC speech and produces language identification as a one-hot vector. (b) The SLIL based ASR network learns the sentence language identification through the FiLM network on the left, which generates gamma and beta values. Then, it helps the encoder learn the language identification in the feature through feature-wise transformation and channel attention mechanism using the SLIL module.}
 \end{figure*}
 
\section{Methodology}\label{sec3}

The multilingual speech recognition framework we propose comprises three key modules: the LID network, the FiLM feature generate network, and the ASR backbone network. The LID network is utilized to acquire sentence language identification (as depicted by the lower left dashed box in Figure 2). Meanwhile, the FiLM network is responsible for learning and integrating sentence language identification into the ASR backbone network (as represented by the lower right dashed box in Figure 2).

To obtain sentence language identification from the input speech, we utilize an RNN-based LID. For the ASR backbone network, RNN-based neural networks can achieve competitive performance even with smaller datasets and fewer computational resources. Therefore, we construct the backbone network of our proposed system with RNN layers and optimize it using the CTC loss function. To embed language identification into the ASR backbone network, we propose a method that employs a SLIL module.

Note that the proposed multilingual recognition framework trains a LID model in the first stage, followed by training an ASR model in the second stage. In this work, we use the trained LID as a component of the system and do not update its parameters in the multilingual ASR training process. In this section, we will introduce the RNN-based LID, the multilingual backbone network, and the proposed SLIL module respectively.

\subsection{LID model}

LID can obtain language identification from speech and provide external language identification as input to augment the input feature for the ASR model, improving the multilingual ASR performance \cite{waters2019leveraging,2018Multi,2018Occam,2019Leveraging,yoo_highly_2019}. Besides being an additional input to the acoustic model, the language identification can also be used in the decoding stage to improve the decoding effect of multilingual speech recognition \cite{zhang_streaming_2022}. Generally, the language identification is a discrete scalar. Motivated by natural language processing, we can denote the language identification with a one-hot vector $L_{i}$. Usually, the LID converts an input speech sequence $X = \{x_{1}, x_{2},\cdot\cdot\cdot, x_{i}\}$ to a frame-level one-hot vector representation sequence $O = \{o_{1}, o_{2}, \cdot\cdot\cdot, o_{i}\}$, where each speech frame corresponds to a one-hot vector of language identification. Moreover, to solve the problem of the varying length of speech samples, we average the acoustic features learned by RNN in the time dimension.

In previous work \cite{2019Leveraging}, the frame-level language identification was used to improve multilingual ASR performance. It improved the recognition of intra-sentence multilingual and code-switch multilingual ASR. For the ATC, both intra-sentence and inter-sentence multilingual recognition problems exist, but inter-sentence multilingual recognition is more common. Motivated by \cite{yoo_highly_2019}, our proposed multilingual speech recognition model uses sentence language identification to improve performance. In this work, as shown in Figure 2 left part, we construct the LID by 3 CNN layers and 5 RNN layers and optimize it by the CE loss function to obtain a one-hot vector representation sequence $O$.

\begin{equation}
  \mathcal{L}_{ce} = -\sum{p(i)\log(q(i))}.
\end{equation}

The CE loss is shown in formula (1), where $p(i)$ is the ground truth and $q(i)$ is the model prediction.

\subsection{SLIL module for multilingual modeling with language identification}
\subsubsection{Appending language identification to speech feature}

Usually, for monolingual speech recognition, the input speech sequence $X$ is the only input of the ASR model. For the model that uses language identification to handle multiple languages, there are two inputs: the speech input sequence $X$ and the external one-hot vector representation $L$ of language identification. Previous works \cite{2018Multi,2018Occam,2019Leveraging,yoo_highly_2019,zhang_streaming_2022} have achieved competitive multilingual speech recognition results by using language identification to enhance speech features. In these works, they either convert the language identification into one-hot vector representations or learn to embed them and append them to the first or each encoder layer. A typical way to enhance features with the one-hot vector of language identification is as follows. Appending a language one-hot vector to layer inputs is equivalent to adding a language-related bias to the input speech feature $X$. The output of layer $i$ neural network is shown in formula (2). We can show the appending process by extending the formula (2).

\begin{equation}
	h_i=f(W\cdot h_{i-1} + b),
\end{equation}

\begin{equation}
\begin{aligned}
	h_{i} & =f(W \cdot[x \mid l]+b) \\
	& =f\left(W_{x} \cdot x+W_{l} \cdot l+b\right).
\end{aligned}
\end{equation}

Where $f(\cdot)$ denotes the activation function, $x$ is the input speech feature and $l$ is the one-hot vector corresponding to $x$. Generally, the last two terms of formula (3), $W_l\cdot l + b$, are independent of $x$ and they are a language-related bias term for the first layer.

\subsubsection{FiLM module}

The previous work \cite{yoo_highly_2019} introduced the FiLM layer for multi-dialect speech recognition. They achieved state-of-the-art results in multi-dialect speech recognition by appending a one-hot vector of dialect information to acoustic features. Unlike previous language identification in multilingual ASR, dialect information is sentence rather than frame-level in this task. However, both types of information can be represented by a one-hot vector or an embedding and appended to the encoder's acoustic feature. We propose the SLIL module that embeds language identification into the encoder layer of the multilingual ASR model to better append sentence dialect information to acoustic features. Following \cite{perez_film_2017} and \cite{yoo_highly_2019}, we use the FiLM mechanism to embed sentence language identification into each frame and channel of the acoustic features. Then we learn special language identification features through the channel attention mechanism.

FiLM affects the output of the network by utilizing a feature-wise transformation of the intermediate features based on some input. More formally, FiLM obtains $(\gamma, \beta)$ from the learns function. In addition, if the sentence language identification $l$ is represented by the one-hot vector of the $D$-dimension where $D$ represents the number of language identification, the $\gamma$ and $\beta$ for all layers can be generated at once as follows:

\begin{equation}
\begin{array}{c}
	a_{c} =\tanh \left(W_{c}\left(\tanh \left(W_{d} d+b_{d}\right)\right)+b_{c}\right), \\
	\cr\left(\gamma^{1}, \ldots \gamma^{L}\right)=\tanh \left(W_{\gamma} a_{c}+b_{\gamma}\right),\\
	\cr\left(\beta^{1}, \ldots \beta^{L}\right)=\tanh \left(W_{\beta} a_{c}+b_{\beta}\right).
\end{array}
\end{equation}
where $ L $ is the total number of encoder layers of the backbone. The $W$s denote weight matrices and the $b$'s are the bias terms. In general, FiLM's extra input (one-hot vector of language identification) through the linear transformation layer and Tanh activation function, get all $\gamma$ and $\beta$.

As shown in Figure 3, the FiLM layers are inserted into the encoder layer of the backbone network, and each such layer applies feature-wise affine transformations to its input as follows:
\begin{equation}
	\hat{\mathbf{x}}=\gamma \odot \mathbf{x}+\beta.
\end{equation}
where $x$ is the FiLM layer input vector, $\hat{\mathbf{x}}$ is the FiLM layer output vector, and $\gamma$ and $\beta$ are the scalings and shifting vectors that are dynamically generated based on an auxiliary input(one-hot vector of language identification), and the $\odot$ denotes the pointwise product of vectors.

\begin{figure}[ht!]
	\includegraphics[width=0.9\linewidth]{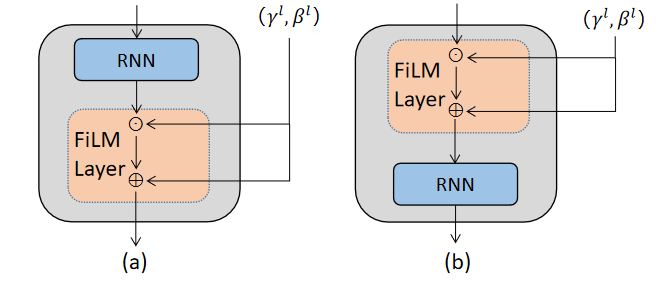}
	\caption{Two different conditioning positions.}
\end{figure}

As shown in Figure 3, the left picture indicates that the position of the FiLM layer is before the encoder layer of the backbone network, and the right picture shows the FiLM layer's position below the encoder layer of the backbone network.

%


\begin{figure}[h]
	\centerline{\includegraphics[width=0.85\linewidth]{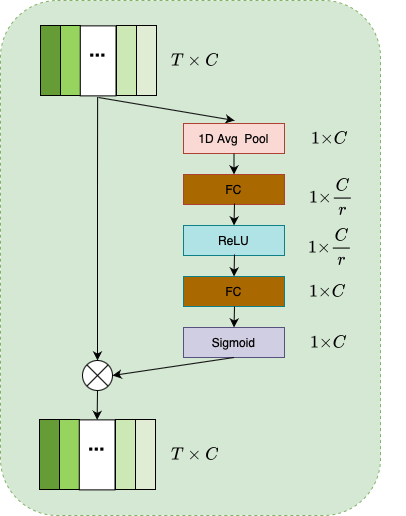}}
	\caption{The 1D Squeeze-and-Excitation module.}
\end{figure}

The SE networks \cite{SE} was introduced in the field of computer vision to improve the feature learning ability of convolutional neural networks and achieved competitive results. SE networks have also been introduced into ASR to enhance the perception field of CNNs, achieving state-of-the-art performance \cite{han2020contextnet}.

The 1D SE network is depicted in Figure 4. The SE network mainly introduces the SE block for channel attention, and the attention mechanism can correct the features, and the corrected features can retain valuable features and eliminate useless features. The SE block is mainly divided into two steps:
\begin{enumerate} 
\item[$\bullet$]The first step is Squeeze: it applies a global average pooling to the input features.
\item[$\bullet$]The second step is Excitation: it adapts and recalibrates the features of the first step by reducing and recovering the dimension and then performing sigmoid activation.

\end{enumerate}
The final operation is to multiply the obtained weight matrix and the input feature to obtain the final feature.
\begin{equation}
\bar{x}=\frac{1}{T} \sum_{t} x_{t}, 
\end{equation}
\begin{equation}
\theta(x) =\operatorname{Sigmoid}\left(W_{2}\left(\operatorname{ReLU}\left(W_{1} \bar{x}+b_{1}\right)\right)+b_{2}\right), \\  
\end{equation}
\begin{equation}
\mathrm{SE}(x) =\theta(x) \circ x .
\end{equation}

In this work, we convert the vanilla 2D SE module into a 1D SE module. Where SE$(\cdot)$ denotes the Squeeze-and-Excitation networks, $x_{t}$ denotes $t$h frame feature, and $\bar{x}$ represents the output after 1D average pooling. The (6) and (7) denote the first squeeze step, and the (8) denotes the second excitation step.


The encoder layer neural network converts an input sequence $X$ of the FBANK to a high-dimensional feature representation:
\begin{equation}
    H = h(X),
\end{equation}

\begin{equation}
    SLIL = SE(FiLM(h(X))).
\end{equation}

In summary, we propose the SLIL module, which first adds sentence language identification to acoustic features through the FiLM layer, and then learns useful feature information through channel attention through the SE layer. As shown in formula (10), this is the proposed SLIL module:
\begin{enumerate}
\item[$\bullet$]The FiLM layer embeds sentence language identification into each frame and channel of the acoustic features.
\item[$\bullet$]The SE layer applies channel attention to the features to enhance the useful ones and suppress the useless ones.
\end{enumerate}

\subsection{FiLMed ASR backbone network}
\subsubsection{The backbone network}
In this work, the backbone network is constructed by cascading the CNN and RNN layers and is optimized with by the CTC loss function. For this end-to-end multilingual speech recognition model, the CNN layer first learns the acoustic features, the RNN layer forms the encoder of the model, and the linear layer and the CTC Loss form the decoder of the model. The batch normalization is applied to speed up the model convergence, while the dropout layer is used to prevent the overfitting problem. The ReLU is selected as the activation function for the proposed model.

\subsubsection{The SLIL module}

As mentioned above, the SLIL module can be inserted into each encoder layer of the backbone network separately. It is optionally inserted before the input acoustic features or after the output acoustic feature of the current encoder layer. Also, to verify the superiority of our proposed SLIL module, we can try to put the SE block before the FiLM block. Therefore, there are four possible ways to insert this module into the backbone network, as shown in Figure 5.

\begin{figure}[h]
	\includegraphics[width=1.05\linewidth]{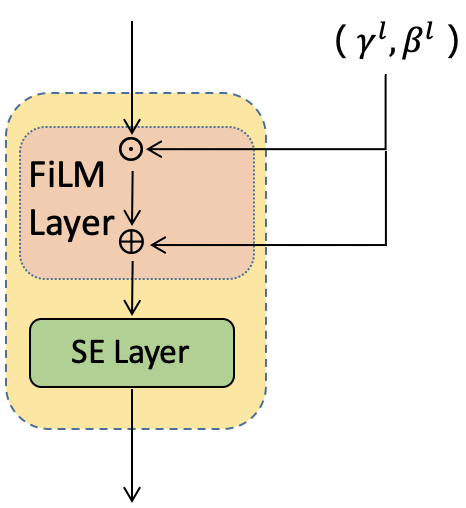}
	\caption{FiLM module and SE network composed of SLIL module and SE-FiLM module. (a) (b) is the SLIL module and the SE-FiLM module perform feature learning after the RNN output features, respectively, and then (c) (d) the SLIL module and the SE-FiLM module perform feature learning before the RNN input features, respectively.}
\end{figure}

\subsubsection{The CTC Loss}

In the proposed end-to-end ASR model, the goal is to predict the text sequence \textcolor{black}{$S = \{s_1,...,s_m\}$ }from the input speech signal \textcolor{black}{$X = \{x_1, ..., x_o\}$}, in which \textcolor{black}{$s_i$} is from a special vocabulary based on Chinese characters and English letters.

In general, multiple frames in $X$ correspond to a token of \textcolor{black}{$S$}. The length of speech frames is usually much longer than the label length. To address this issue, the CTC loss function was designed to automatically achieve the alignment between the speech and label sequence. The $t$th frame corresponds to the output label $k$ and its probability is denoted \textcolor{black}{$z_{\pi_t}^t$}. Given the speech input $X$, the probability of the output sequence $\pi$ is shown in (11). Therefore, the probability of the final sequence can be obtained by (12), in which \textcolor{black}{$v$} is the set of all possible sequences and \textcolor{black}{$A$ denote the length $T$ sequences
over the vocabulary}. For example, by using '$\_$' to denote a blank, both the outputs "$X\_YY\_Z$" and "${\_XY\_Z\_}$" correspond to the final output "$XYZ$" \cite{graves2006connectionist}.

\begin{equation}
\textcolor{black}{p(\pi|X)=\prod_{t=1}^T{z_{\pi_t}^t}, \pi\in A.}
\end{equation}
\begin{equation}
\textcolor{black}{p(S|X)=\sum_{\pi\in v^{-1}(S)}p(\pi|X).}
\end{equation}

\section{Experiments}\label{sec4}
\subsection{Datasets}

In this work, the training data of the proposed model is the ATCSpeech corpus, which is collected from the real-world ATC environment, and manually annotated \cite{yang_atcspeech_2020}. The ATCSpeech corpus is a multilingual corpus that contains both Chinese and English speeches in different flight phases and areas. The corpus has inter-sentence multilingual speech (separate sentences in different languages) and intra-sentence multilingual speech (mixed sentences with both languages). The corpus has 16111 English utterances (about 17.48-hour) and 28927 Chinese utterances (about 23.23-hour) and 16645 mixed Chinese and English utterances (about 15.81-hour), all with the 8000 Hz sample rate. Table 1 shows the division for the train, validation, and test set.

\begin{table}[h]
\renewcommand\arraystretch{1}
\label{tab0}
\caption{Data size of the corpus. "\#U" denotes the speeches utterances and "\#H" denotes the speeches hours\textcolor{black}{.}}
\setlength{\tabcolsep}{2.0mm}
\begin{tabular}{cccccccc}
\hline
\hline
\multirow{2}{*}{\textbf{Language}} & 
\multicolumn{2}{c}{Train} & \multicolumn{2}{c}{Dev} & \multicolumn{2}{c}{Test} \\ \cline{2-7} 
&\#U & \#H & \#U & \#H & \#U &\#H\ \\
\hline
Chinese & 27391 & 21.99 & 766 & 0.62 & 770 & 0.62 \\
English &	15282 &	16.58 & 425 & 0.46 & 404 & 0.44\\
Mixed & 15781 & 15.00 & 434 &0.41 &430 &0.40\\
All &	58454 &	53.57 & 1625 & 1.49 & 1604 & 1.46\\

\hline
\end{tabular}
\end{table}

\subsection{Experimental settings}

In this work, the proposed model is constructed based on the open framework PyTorch 1.7.0. The training server was equipped with an Intel i7-9700 processor, a single NVIDIA TITAN RTX GPU, 32-GB memory, and an Ubuntu 18.04 operating system.

During the model training, the Adam optimizer is used to optimize the trainable parameters. The initial learning rate is 0.0001. The batch size is set to 32. In the first epoch, the speech samples are sorted in reverse order (based on speech duration) to detect the overflow of GPU memory as early as possible. In the following epochs, the training samples are shuffled to improve the model's robustness. The vocabulary is built on Chinese characters and English letters, and also with some special tokens ($< unk >$, $< blank >$). Finally, a total of 682 tokens in the vocabulary.  In this work, the Deep Speech 2 \cite{pmlr-v48-amodei16}, Jasper \cite{li2019jasper} and Conformer \cite{gulati2020conformer}, which are applied to achieve the multilingual ASR task in this work. In order to ensure the fairness of the experiment, all those models are trained on the same dataset (ATCSpeech) without extra training data.

\subsection{Evaluation metrics}
In this paper, the output of the model contains both Chinese characters and English words. Words Error Rate (WER) is not very suitable for evaluation, so we choose CER to evaluate the Chinese characters and English letters in the sentences by the model output. 
 
\begin{equation}
    CER=\frac{S+D+I}{N}.
\end{equation}

The CER is the conventional ASR evaluation metric, which describes the difference between the target token sequence and predictions as Eq(8). Where $N$ is the length of the target token sequence, the $S$, $D$, and $I$ are the number of the substitution, delete and insert operations for converting the predictions into the target token sequence.

\subsection{Overall Results}

\begin{table}[h]
\caption{The overall result of the multilingual speech.}
\label{tab:my-table}
\begin{center}
\setlength{\tabcolsep}{5.2mm}
\begin{tabular}{ccc}

\hline
\hline
\multirow{2}{*}{Methods}                        & Dev     & Test    \\ \cline{2-3} 
                                             & CER(\%) & CER(\%) \\ \hline
Deep Speech 2 \cite{pmlr-v48-amodei16} & 7.01    & 6.98    \\
            
Conformer \cite{gulati2020conformer}    &  7.65    & 7.66   \\
          
Jasper10*3 \cite{li2019jasper}   & 8.61   & 8.59    \\
\hline
FiLMed network         &     5.71     & 5.73   \\

The proposed Method        & \textbf{  5.28 }     & \textbf{ 5.30 } \\
\hline
\end{tabular}
\end{center}
\end{table}

In this work, we compare the results without external language model to fairly compare the true representation power of the model architectures alone.

We applied all the models to the multilingual ASR task in the ATC ASR task. Our LID model achieved 99.95 \% accuracy, which means that the language classification results had little impact on the multilingual speech recognition model. Table 2 shows the performance of three multilingual speech recognition models without language identification on ATCSpeech as baselines. Among them, the Deep Speech 2 model performed better than the others. The conformer model did not perform well due to limited dataset size and batch size. It achieved 6.98\% CER in the test set, which was 0.68\% lower than the self-attention based Conformer model and 1.61\% lower than the CNN based Jasper10*3 model. Therefore, we chose the Deep Speech 2 model as the backbone network of our work.

Table 2 shows the performance comparison between the method proposed in this article, which utilizes sentence language identification directly with the FiLM module and the baselines. The FiLMed network achieved CER of 5.73\% on the test sets, outperforming the optimal baselines (Deep Speech 2) by reducing 1.28\%. Moreover, the proposed method achieved CER of 5.30\% on the test sets, surpassing the FiLMed network by 0.43\%. Overall, both methods utilized sentence language identification to improve multilingual speech recognition performance beyond the best baseline (Deep Speech 2), but the proposed method demonstrated superior performance due to its better ability to leverage sentence language identification.

\subsection{Ablation Studies}

\begin{table}[h]
\caption{The results of the SE network at distinct locations within the proposed module, and the proposed module at varying conditional positions in the backbone network.}
\label{tab:my-table}

\setlength{\tabcolsep}{4.2mm}
\begin{tabular}{cccc}
\hline
\hline
\multirow{2}{*}{Methods}          & Language               & Dev     & Test    \\ \cline{3-4} 
                                  & information            & CER(\%) & CER(\%) \\ \hline

M1         & Before                                       &    5.71     & 5.73   \\
M2         & After                                     &    5.95     &  5.94  \\
\hline
M3         & Before                                   &  \textbf{  5.28 }     & \textbf{ 5.30 } \\
M4         & Before                                   &    5.45     & 5.44   \\
M5         & After                                    &    5.51     &  5.50  \\
M6         & After                                    &    5.63     &  5.65  \\

\hline

\end{tabular}
\end{table}

We designed several experiments to validate the multilingual ASR model based on the proposed method. We used Deep Speech 2 as the backbone network and applied FiLM to embed language identification into the encoder layers. Table 3 summarizes the details of the experiments. We trained six ASR models on ATCSpeech with different configurations of FiLM and SE layers, as follows: 
\par a) M1: FiLM layer before encoder layer; 
\par b) M2: FiLM layer after encoder layer; 
\par c) M3: SLIL layer before encoder layer; 
\par d) M4: SE-FiLM layer before encoder layer; 
\par e) M5: SLIL layer after encoder layer; 
\par f) M6: SE-FiLM layer after encoder layer.

M1 and M2 are models that take feature-wise transformations based on language identification to each encoder layer. The difference is that M1 applies them before each encoder layer and M2 applies them after. M1 achieved 5.73\% CER on the test set, which was 0.21\% lower than M2. Both models with FiLM added language identification performed better than the vanilla Deep Speech 2 ASR model (6.98\% CER). This improvement suggests that the FiLM module helps the encoder use language identification to enhance the multilingual speech recognition performance of the model.


M3 and M5 models use the SLIL module, which consists of a feature-wise transformation followed by a channel attention layer. The difference is that M3 applies the SLIL module before each encoder layer and M5 applies it after. M3 achieved 5.30\% CER on the test set, which was 0.20\% lower than M5. This result is similar to the comparison between M1 and M2, it shows that adding language identification after the encoder layer can make the model perform better. The SE-FiLM module in M4 and M6 performs channel attention on the acoustic features before the feature-wise transformation. The difference is that M4 applies the SE-FiLM module before each encoder layer and M6 applies it after. M4 achieved 5.44\% CER on the test set, which was 0.21\% lower than M6. When comparing M3 and M4, M3 performed better by 0.14\%. This may be because the SLIL module embeds language identification into the features first and then uses channel attention to retain the language-related channels and suppress the irrelevant ones. This helps the AM learn language identification better.

\begin{table}[h]
\caption{The result of the SLIL with different backbone network}
\label{tab:my-table}
\begin{center}
\setlength{\tabcolsep}{2.2mm}
\begin{tabular}{ccc}

\hline
\hline
\multirow{2}{*}{Backbone network}                        & Dev     & Test    \\ \cline{2-3} 
                                             & CER(\%) & CER(\%) \\ \hline
Deep Speech 2 \cite{pmlr-v48-amodei16} & 7.01    & 6.98    \\
            
Conformer \cite{gulati2020conformer}    &  7.65    & 7.66   \\
\hline
                                             
SLIL with Deep Speech 2  & \textbf{5.28}    & \textbf{5.30}    \\
            
SLIL with Conformer     &  5.97    & 5.96   \\
          
\hline
\end{tabular}
\end{center}
\end{table}
To assess the effectiveness of the SLIL proposed in this paper, we performed multilingual speech recognition experiments on ATCSpeech, leveraging both Deep Speech 2 and Conformer as the backbone networks, and combining them with the SLIL method. Table 4 displays the experimental results, demonstrating that both new models incorporating SLIL outperformed their vanilla counterparts. These findings support the efficacy of our proposed method in enhancing the multilingual speech recognition performance of the model.

\section{Conclusion}\label{sec5}

In this work, we propose a two-stage multilingual ASR framework with a SLIL module for the intelligent ATC system. We introduce the FiLM module into multilingual ASR in this study, utilizing sentence language identification to enhance the performance of multilingual ASR models. Furthermore, we proposed a SLIL module construct of FiLM and SE modules. The proposed method has two-stage. The first stage is to train an RNN-based LID that obtains sentence language identification from ATCSpeech. The second stage is to utilize sentence language identification to augment the input feature for training a powerful end-to-end multilingual ASR model. The results show that our method achieves state-of-the-art ASR performance on ATCSpeech. The ablation studies confirm that all the components in our method are effective.

\section{Acknowledgments}

The authors would like to thank Sining Sun for his help on refining our manuscript. This work was supported by the National Natural Science Foundation of China under Grants 62001315 and U20A20161, the Open Fund of Key Laboratory of Flight Techniques and Flight Safety, Civil Aviation Administration of China (CAAC) under Grant No. FZ2021KF04, and
Fundamental Research Funds for the Central Universities under Grant No. 2021SCU12050. The authors would like to thank all contributors to the ATCSpeech corpus.

\bibliography{main_bib}

\end{document}